# A SURVEY ON COOPERATIVE DIVERSITY AND ITS APPLICATIONS IN VARIOUS WIRELESS NETWORKS

Gurpreet Kaur[1] and Partha Pratim Bhattacharya[2]

Department of Electronics and Communication Engineering
Faculty of Engineering and Technology
Mody Institute of Technology & Science (Deemed university)
Lakshmangarh, Dist. Sikar, Rajasthan,
Pin – 332311, India

[1]gur1487preet@gmail.com
[2]hereispartha@gmail.com

## ABSTRACT

*Cooperative diversity is a technique in which various radio terminals relay signals for each other. Cooperative diversity results when cooperative communications is used primarily to leverage the spatial diversity available among distributed radios. In this paper different cooperative diversity schemes and their applications in various wireless networks are discussed. In this paper the impact of cooperative diversity on the energy consumption and lifetime of sensor network and the impact of cooperation in cognitive radio are discussed. Here, user scheduling and radio resource allocation techniques are also discussed which are developed in order to efficiently integrate various cooperative diversity schemes for the emerging IEEE 802.16j based systems.*

## 1. INTRODUCTION TO COOPERATIVE DIVERSITY

Cooperative diversity is a diversity technique which is obtained when relay nodes are used for transmitting the signals. The source node transmits two independent signals to the relay node and the destination node, and the destination node receives signal from the source and the retransmitted signal from the relay node. With the help of relaying node the quality of the signal received at the destination can be improved. For notational simplification, system with three nodes namely source, relay and destination is considered.

The history of the cooperative communication can find its deep roots to the groundbreaking work of Van der Meulen [1], when he introduced the concept of relay channel model, the channel model consists of a source, destination and relay; and whose major purpose was to facilitate the information transfer from source to destination. Later, Cover and El Gamal [2] deeply investigated the relay channel model, which provided a number of fundamental relaying techniques such as Decode and Forward (DF) and Compress and Forward (CF). In conventional communication, data is transmitted between the source and destination, and no user provides assistance to one another demonstrated in Figure 1.

There are many neighboring nodes in a practical wireless communication network, which could be of great assistance. When one node transmits its data, all the nearby nodes overheard its transmission. Cooperative Communication aims to process and forward this overhead information to the respective destination to create spatial diversity, which results in increasing the system performance. The concept of the cooperative communication is suggested in Figure 2 [3].

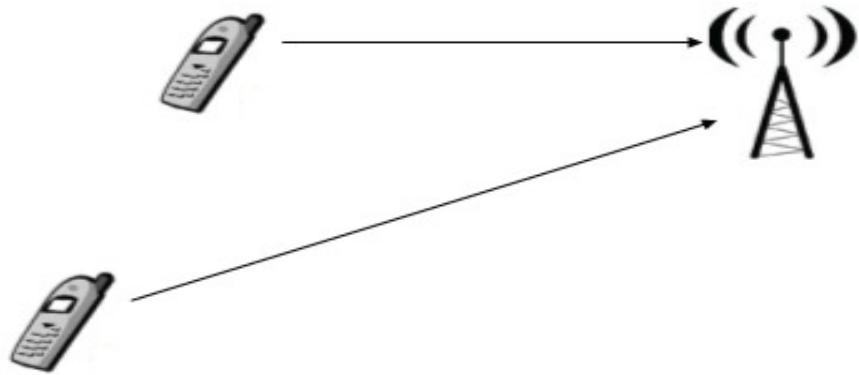

Figure 1.Conventional communication

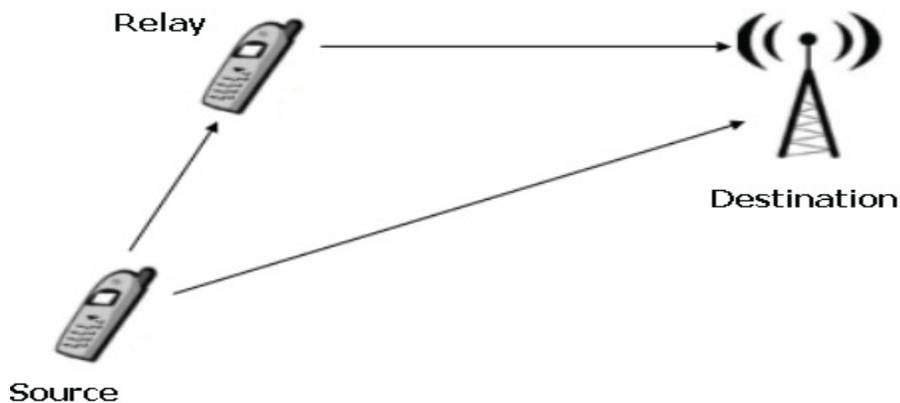

Figure 2.Cooperative communication

As shown in Figure 2, the source node is transmitting the data to the destination node; while the relay node (another mobile user) is also helping in the transmission. The relay station also process and forward this message to the destination, where both of the received signals are combined. As both of the signals are transmitted through independent paths, this results into spatial diversity. In cooperative communication, each wireless user is assumed to transmit its own data as well as act as a cooperative agent (relay) for the other user.

## 2. COOPERATIVE TRANSMISSION PROTOCOLS

The processing of the signal at the relay node which is received from the source is described with the help of cooperative transmission protocols. Different transmission protocols are discussed here.

### 2.1. Decode and forward

The most popular method for processing the signal at the relay node is decode and forward, in this technique, the relay detects the source data, decodes and then transmits it to the desired destination. The concept of the Decode and Forward technique is shown in Figure 3.

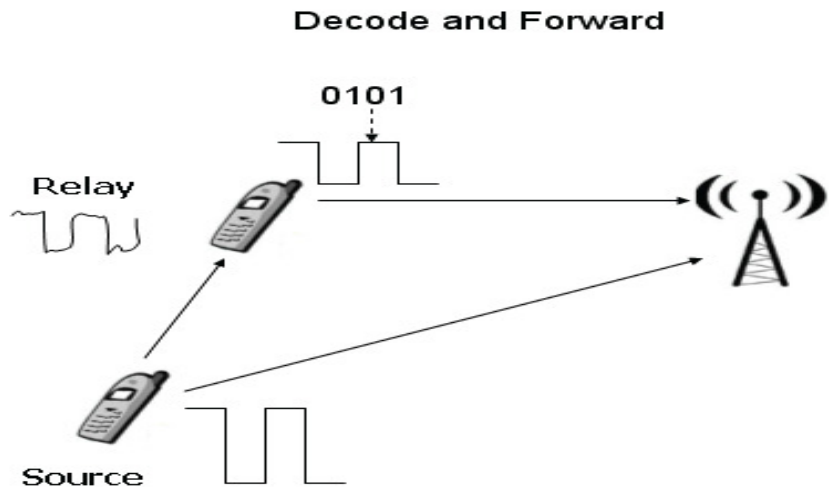

Figure 3.Decode and Forward technique

An error correcting code can also be implemented at the relay station. This can help the received bit errors to be corrected at the relay station. However, this is only possible, if the relay station has enough computing power [4].

### 2.2. Amplify and forward

Amplify and Forward technique simply amplifies the signal received by the relay before forwarding it to the destination. This technique was proposed by J. N. Laneman and G. W. Wornell [5], and is ideal when the relay station has minimal computing power. However, one major drawback of this technique is that the noise in the signal is also amplified at the relay station, and the destination receives two independently faded versions of the signal. Figure 4 shows amplify and forward technique.

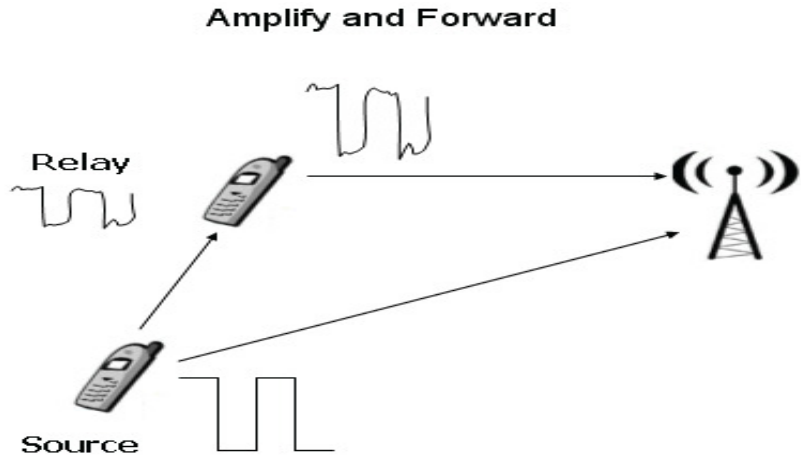

Figure 4.Amplify and forward technique

## 3. APPLICATIONS OF COOPERATIVE DIVERSITY

Cooperative diversity can be used in various fields like cooperative sensing in cognitive radio, wireless ad hoc networks, wireless sensor networks and many more. Different applications of cooperative diversity are described in the subsequent sections.

### 3.1. Cooperative diversity in wireless sensor networks

Wireless sensor networks (WSNs) are a broad class of wireless networks consisting of small, inexpensive and energy limited devices [6]. Due to the fact that nodes are battery powered, energy efficiency is one of the main challenges in designing Wireless sensor networks. Schemes have been developed recently for energy saving of the protocol stack in specific layers. For example, multi-hop routing and clustering improve the energy efficiency of large scale WSNs. As nodes can communicate directly over small distances and have limited transmission range multi hop routing is necessary. However, it is restricted to networks of extremely high densities [7]. Clustering is a method of partitioning the network into local clusters, and each cluster has a node called cluster-head (CH).

Energy saving protocols has also been developed in the physical layer. Like all other wireless networks, wireless sensor networks suffer from the effects of fading. Cooperative diversity is a technique used to mitigate the impact of fading. This form of diversity is especially suited towards WSNs since size and power constraints restrict nodes from possessing more than one antenna. Cooperation is achieved using the simple amplify-and-forward scheme [8].These results can be used to predict the impact of cooperative diversity on the lifetime of sensor networks. Here different design aspects of cooperative diversity used in wireless sensor network are discussed.

### 3.1.1. Clustering Protocol

The network is clustered using a distributed algorithm where CHs are selected randomly. These classes of algorithms are practical to implement in WSNs since WSNs are organized in a distributed fashion. The role of CH is evenly distributed over the network and each CH performs ideal aggregation, i.e., all cluster data is aggregated into a single packet.

### 3.1.2. Routing Protocol

Multi-hop path from each Cluster head to the data sink is established with the help of multi hop routing (MHR). MHR gives a good performance in stationary networks comprising of nodes having fixed transmission power levels. A simple iterative algorithm is used in MHR that begins with broadcasting the nodal hop number for the nodes neighboring the data sink. The neighboring nodes in turn update and broadcast their hop number and the process continues until each node in the network determines its min-hop path to the data sink.

### 3.1.3. Cooperative Diversity

Each sensor in the multi-hop path has a cooperative partner and each hop is no different than the three node network studied in [8]. Figure 5 illustrates one section of a multi-hop path where nodes M1 and node M2 belong to the multi-hop path and node C represents a potential cooperative relay. All channels are modeled as slow and flat. The receiver is assumed to know the channel perfectly.

The cooperating node, C in Figure 5, helps in the communication between nodes M1 and M2 using the amplify-and-forward (AF) protocol. Thus node C receives a noisy version of M1's transmitted signal and transmits an amplified version of this signal to M2. This protocol creates spatial diversity since node M2 receives two independently faded signals. The quality of channel between source and relay and that between relay and destination examines the performance of amplify and forward protocol. Since generally channel quality decreases with distance, one should restrict the selection of relays to a node's forward transmission region.

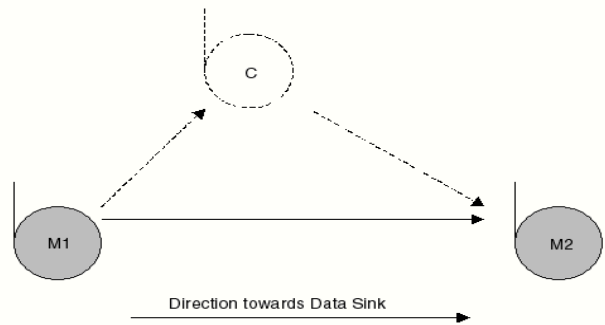

Figure 5.

### 3.1.4. Analysis of multi-hop routing

The transmission of packets dominates the energy consumption of sensors; hence this analysis represents the first step towards predicting the energy consumption of the network. Low-to-medium density networks are considered here. In this analysis each layer has a width $R_1$ (corresponding to the communication radius used by nodes for multi-hop transmission). Let r denote the distance of a node from the data sink and differentiate our framework by allowing nodes to forward packets within their own layer, thereby approximating MHR at much lower node densities. As shown in Figure 6(a), for a given node x, nodes that may potentially forward packets to x lie in a circle of radius $R_1$ centered at x. Since nodes are assumed to transmit forward towards the data sink only, x can only receive packets from nodes in the shaded region of Figure 6(a). The layer structure allows for differentiation of the routing behavior of nodes in this shaded region based on whether they are located in the same layer as x.

If a willing multi-hop partner is available in a higher layer, a node preferentially forwards its data to the higher layer. For example, consider Figure 6(b), where node x and node b are in different layers. Routing behavior of b is indifferent to forwarding packets to x and any other node in the shaded area of Figure 6(b). Lastly consider the case where x and b are located in the same layer and the routing behavior of b is such so as to prefer forwarding packets to the shaded region in Figure 6(c). Consequently if no nodes exist in this region, b is modeled to be indifferent to forwarding packets to node x and any other node in the area of intersection of b's transmission region and b's layer.

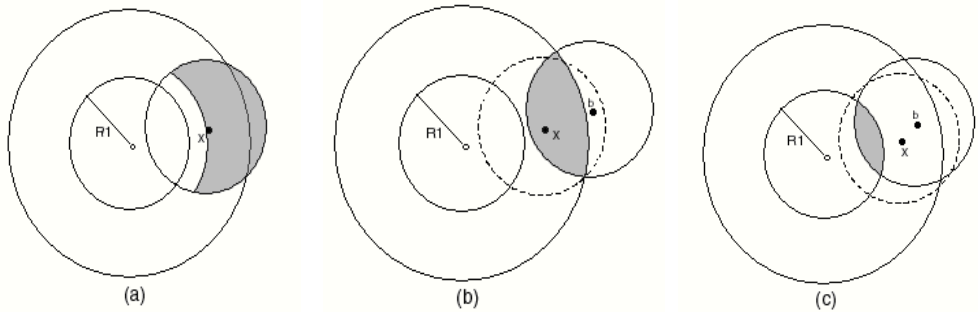

Figure 6. Example topology for multi-hop routing

To determine N (|x|), the expected number of packets forwarded by a node at distance |x| from the data sink, recursively integrate over the shaded region in Figure 6(a) and add one to account for the packet originating at x. Thus N (|x|) is expressed

$$N(|x|) = 1 + \lambda \int_{|x|}^{|x|+R_1} p(b,x)N(|b|)2\gamma|b|\,dr, \qquad (1)$$

where $\lambda$ is the spatial density of nodes and

$$\gamma = \arccos \frac{|x|^2 + |b|^2 - R_1^2}{2|b||x|} \qquad (2)$$

Thus for the case when $|x| \leq R_1$, replace |x| the lower limit for the integration in (1), with $R_1$.

Figure 7 compares the number of packets forwarded versus distance determined theoretically using equations (1) and (2) with simulations based on MHR. The simulations average over 200 different networks at a density of

$$30/(\pi R_1^2),$$

where $R_1 = 1$ and network radius $a = 4R_1$. Clearly the theoretical analysis compares fairly well to the simulations, especially in the first layer, the most critical layer in the network.

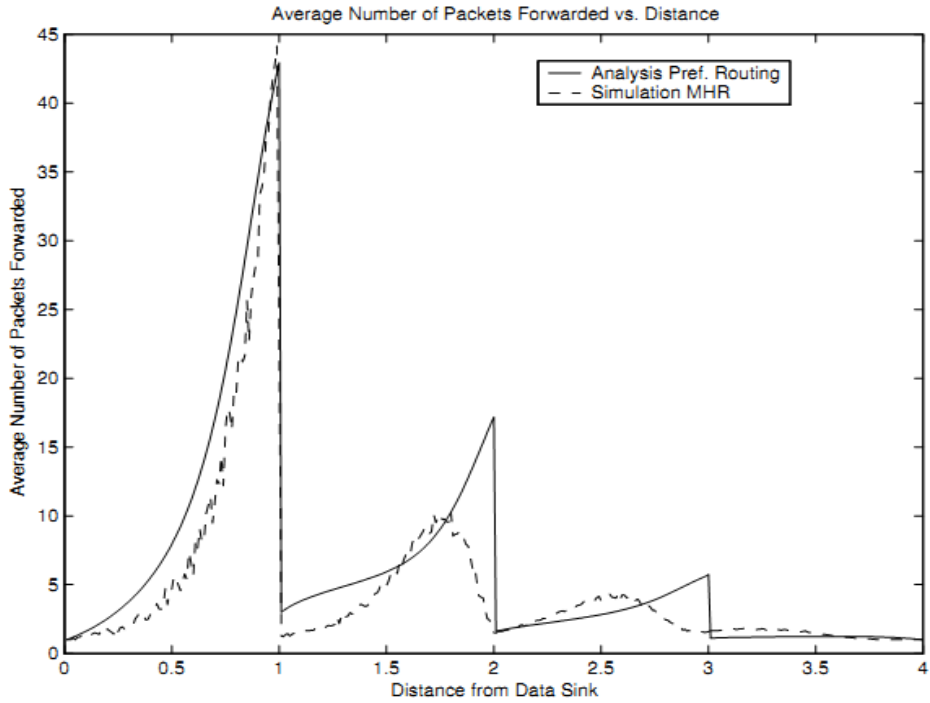

Figure 7.Expected number of packets forwarded versus distance

### 3.1.5. Analysis of cooperative relay selection

The behavior of cooperative relay selection is observed in order to determine the number of cooperative packets forwarded as a function of distance. The performance of the AF protocol depends on the position of the relay relative to the source and destination. Assume node requests cooperation only from other nodes in its forward transmission region with the corresponding radius as R2. This ensures that the relay is relatively close to both the source and destination.

Determine C (|x|), the expected number of packets received exclusively due to cooperation by a node at distance |x| from the data sink.

Thus C (|x|) is expressed as

$$C(|x|) = \lambda \int_{|x|}^{|x|+R2} p_c\ (b,x) N(|b|) 2\gamma |b| \, d x, \qquad (3)$$

Where

$$\gamma = \arccos \frac{|x|^2 + |b|^2 - R_2^2}{2|b||x|}, \qquad (4)$$

where N (|b |) is the number of packets forwarded. Figure 8 is the counterpart of Figure 7 for networks with cooperation. Again the simulations average over 200 different networks at a density of

$$30/(\pi R_1^2),$$

where $R_1 = 1$ and network radius $a = 4R_1$. It is observed from Figure 8 that (3) slightly overestimates the number of cooperative packets forwarded. This is due to the dependence on preferential routing to determine N (lb l). The analysis in (1) and (3) illustrates the need to assume preferential routing and that packets are transmitted forward. Without these assumptions the simple recursion in these equations is invalid, requiring a complicated iterative scheme, thereby making the theory largely useless.

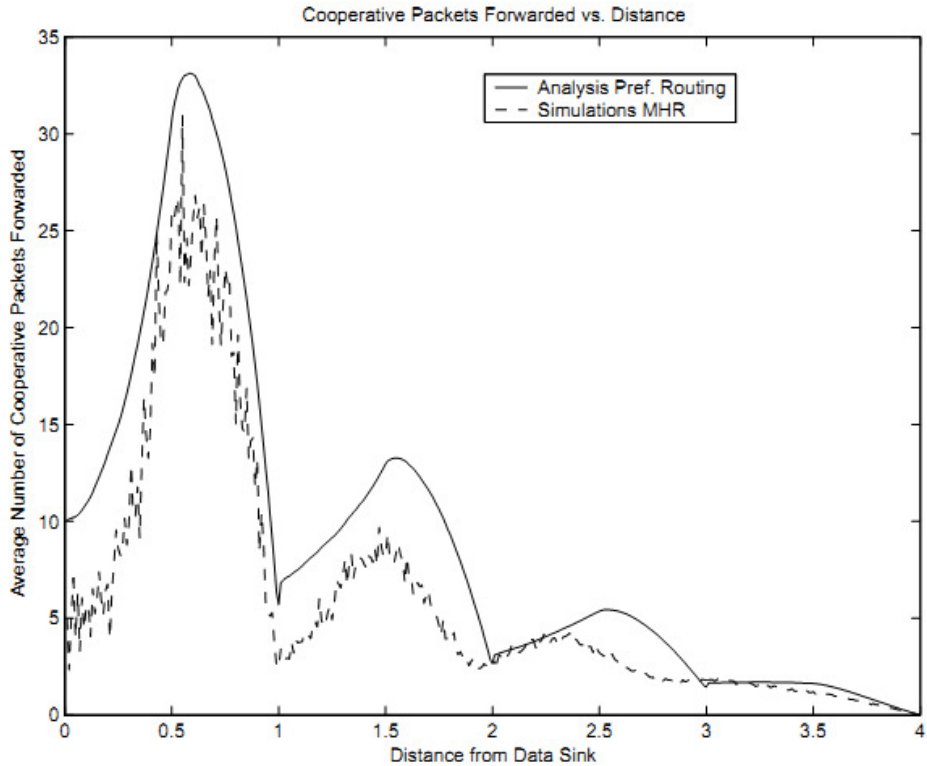

Figure 8.Expected number of cooperative packets forwarded versus distance

## 3.2. Cooperative diversity in cognitive radio

In Software defined radio (SDR), the software embedded in a radio cell phone defines the parameters under which the phone should operate in real-time as its user moves from one place to another. Cognitive radio (CR) is a smarter technology. Cognitive radio is a radio that is meant to be aware, sense and learn from its environment and to serve best to its user. The cognitive users are required to detect the presence of licensed (primary) users in a very short time and must vacate the band for use by primary users. Thus the main challenge in this technology is how to detect the presence of primary users.

Hence diversity gain is achieved by allowing the users to cooperate. Cooperative schemes in a TDMA system with orthogonal transmission have been recently proposed in [9] and [10]. Here different design aspects of cooperative diversity used in cognitive radio are discussed.

### 3.2.1. Problem formulation

All users are assumed to experience Rayleigh fading which is independent from user to user. If a signal sent at the transmitter is *x*, the received signal *y* is given by *y = hx + w,* where *h* is the fading coefficient modelled as a complex Gaussian random variable and *w* is additive white Gaussian noise. Unless otherwise mentioned all noise coefficients are assumed to be with zero-mean and unit-variance. The main requirement of cognitive radio architecture is to detect the presence of the primary or licensed users as quickly as possible. For this reason, the cognitive users should continuously monitor the spectrum for the primary (licensed) users. In Figure 9, two cognitive radio users U1 and U2 are shown to be operating in a fixed TDMA mode in order to send data to a common receiver and the primary user starts using the band as soon as the data is sent. The two cognitive users are then required to vacate the band as soon as possible so that the primary user can use it. The signal received from the primary user is so weak that the cognitive user U1 takes a lot of time to sense its presence.

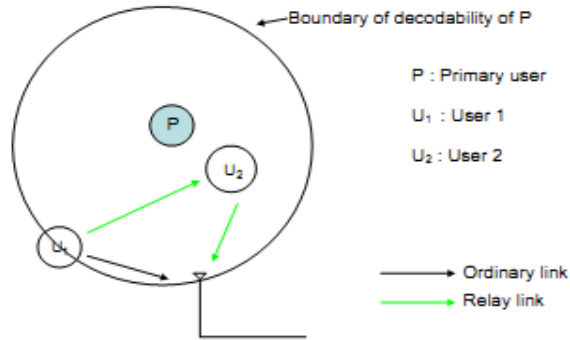

Figure 9.Cooperation in cognitive network

Figure 10 illustrates the protocol used by U1 and U2 to transmit data to some common receiver. A slotted transmission system is used where U1 and U2 transmit in successive slots according to the AF protocol in the same frequency band. Accordingly in time slot T1, U1 transmits and in T2, U2 relays the information of the previous slot. A primary user starts using the band without informing any of the two users.

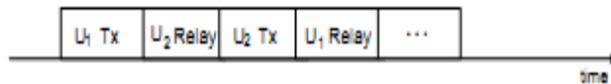

Figure 10.Relay protocol used

Also consider an energy detector(ED) so as to show the effect of cooperation in cognitive networks. The signal is modeled as a random variable with known power.

Hence ED is optimal. If given that $h_{12} = h_{21}$, the random variables H and W in are Rayleigh distributed with zero-mean and variances.

The ED forms the statistics

$$T\,(Y) = |Y|^2 \qquad\qquad (5)$$

and is compared with threshold λ which is determined by some pre specified probability of false alarm α.

In Figure 11 performance curve of the energy detector with and without cooperation is shown. Simulation requires an assumption of $P_1$ =1 and $P_2$ =2.7. The relay user is supposed to have no power constraint. In cooperation technique, the relay terminal (irrespective of the position of U2) so adjusts it's transmit power so that the channel gain between the users is kept constant. This is an important point that affects the performance of cooperation schemes. From the graph it can be seen that we have higher detection probability through cooperation for same probability of false alarm. This is the effect of cooperation in the network.

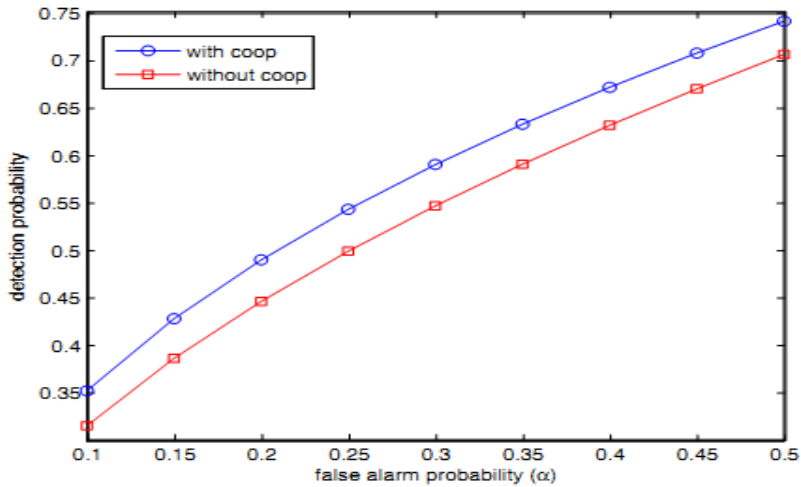

Figure 11.Performance curves of the energy detector with and without cooperation

### 3.2.2. Agility of the two user cognitive radio network

Under the cooperation scheme described above, the average detection time is reduced thus implying an increase in agility. Let $\tau_n$ be the number of slots taken by user U1 in a non-cooperative network to detect the presence of the primary user. This detection time $\tau_n$ can be modeled as a geometric random variable, i.e.

$$P_r\{\tau_n = k\} = (1 - p_n^{(1)})^{k-1} p_n^{(1)}, \tag{6}$$

where $p_n^{(i)}$ denotes the probability of detection by user Ui in a single slot under the non-cooperation scheme. For the system model,

$$p_n^{(1)} = \alpha^{\frac{1}{P_1+1}} \tag{7} \quad \text{and}$$

$$p_n^{(2)} = \alpha^{\frac{1}{P_2+1}} \tag{8}.$$

We define the agility gain of our cooperation scheme over the non-cooperation scheme when there are two users as

$$\mu_{\frac{n}{c}}(2) = \frac{T_n}{T_c} \qquad (9).$$

Note that the agility gain $\mu$ (2) is a function of P1 and P2. However, without loss of generality we set P1 =1 and consider the agility gain as a function of P2 alone. It can be shown that as P2 →∞, then the asymptotic agility gain is given by,

$$\mu_{\frac{n}{c}}^{\infty}(2) = \lim_{P_2 \to \infty} \mu_{\frac{n}{c}}(2) = \frac{1}{\sqrt{\alpha}} \qquad (10).$$

The non-cooperation scheme presented above assumes extremely selfish users. Hence cognitive user U1 detects the presence of the primary user without the help of U2. Also once detected U1 vacates the band without informing U2. Consider now a more practical situation where both the users get informed once either U1 or U2 detects the primary user. Also the detection information is transmitted to the common receiver. The common receiver then informs the other cognitive users. Note that this is a partially cooperative network where the cognitive users still detect the primary user without any cooperation. In such a case, the time taken to vacate the band is given by,

$$T_p = \frac{2 - \frac{P_n^{(1)} + P_n^{(2)}}{2}}{P_n^{(1)} + P_n^{(2)} - P_n^{(1)} \cdot P_n^{(2)}} \qquad (11).$$

The asymptotic agility gain in this case can be shown to be,

$$\mu_{\frac{p}{c}}^{\infty}(2) = \lim_{P_2 \to \infty} \frac{T_p}{T_c} = \frac{4}{3\sqrt{\alpha} - \alpha} \qquad (12).$$

In Figure 12, plotted is the agility gain $\mu_{n/c}(2)$ of a two user asymmetric network as a function of the asymmetry $P_2$ for different values of false alarm probability $\alpha$. As the network becomes more and more asymmetric, agility gain increases.

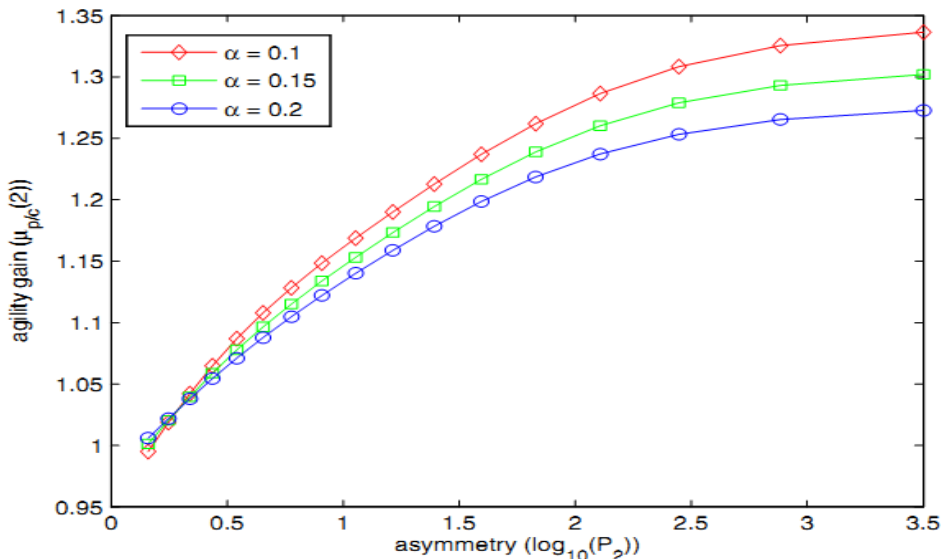

Figure 12. Agility gain in single carrier two user asymmetric network.

### 3.2.3. Agility of the multiuser cognitive network

Let that there are 2n cognitive users. The total bandwidth B is equally divided into n sub-bands each of bandwidth $\Delta_n = B/n$. There are two cognitive users working on each sub-band following the cooperation protocol. It must be noted that the primary user, if present, uses the whole bandwidth B. If P1 denotes the power received from the primary user at cognitive user $U_1$ when it uses the whole band, the signal power received at $U_1$ now is given by $P_1{}' = P_1/n$. The noise power at $U_1$ is similarly scaled by a factor of n. suppose that users $U_1$ and $U_2$ form a link in a particular sub-band with $U_2$ acting as a relay for $U_1$.

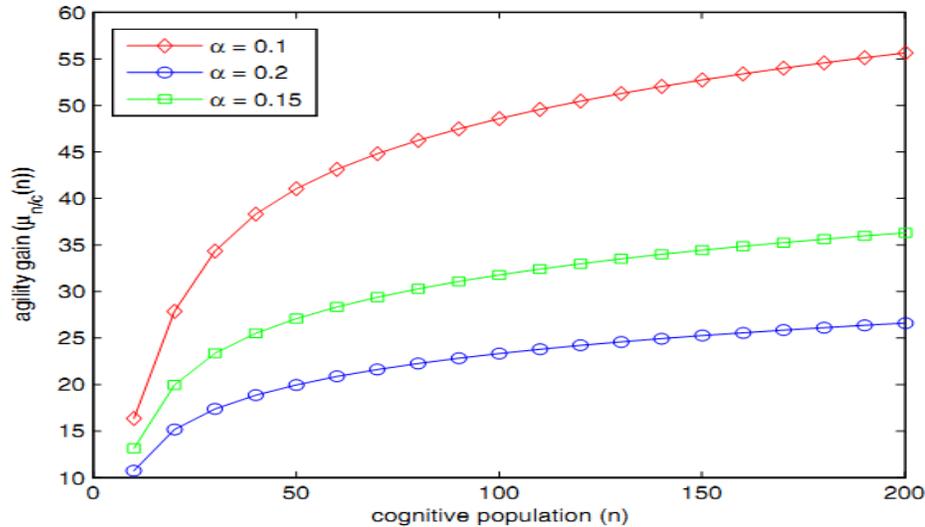

Figure 13. Agility gain in multi-carrier multi-user symmetric networks

## 3.3. Efficient cooperative diversity schemes for IEEE 802.16j

The IEEE 801.16e standard which is based on Orthogonal Frequency Division Multiple Access (OFDMA) provides high data rate to the mobile users in a cell with an approximate coverage radius of 8 km [11]. Developments are made in the form of IEEE 802.16j for increasing the coverage area of IEEE 802.16e standard.

The wireless terminals have a drawback of not being able to transmit and receive simultaneously at the same time, relaying requires two phases. First phase involves source to relay communication and in the second phase information is transmitted from relay to destination.

Multiplexing loss occurs because of the two phase communication since each data block is transmitted twice. Thus scheduling and radio resource allocation needs modifications in the case of multi hop cellular networks over conventional scheduling algorithms designed for the single-hop networks. This is due to the fact that apart from signal to noise ratio the end-to-end performance which also includes the effect of multiplexing loss should be considered. Multi-hop cooperation schemes should only be used only when it has the ability to provide end-to-end throughput greater than that of direct transmission (without relay). In a practical multi-user scenario the work on performances of cooperative diversity schemes is limited. In [12], it can be seen that when relay node can correctly decode the packets from source, the cooperative relay transmissions are used, which in turn causes throughput loss. In [13], end-to-end link adaptation and link selection methods have been developed for a single user in an Orthogonal Frequency

Division Multiplexing (OFDM) Time Division Duplex (TDD) based wireless relay network. The emerging IEEE 802.16j standard may allow without relay transmissions in the second phase. However the amount of radio resource allocation which will be done is not specified by the current standard. It is too complex to do the radio resource allocation along with path selection for each sub-channel. The different design aspects for IEEE 802.16j using Cooperative diversity are discussed below.

### 3.3.1. System model

IEEE 802.16j based on two-hop cellular network is considered here. The without relay system corresponds to the single hop IEEE 802.16e based cellular network. Multiple users and multiple (fixed) relays are considered within a single cell. Low mobility users are considered. Thus during one frame the channel gain of each sub-channel remain unchanged. A sub-channel comprises of multiple sub-carriers with approximately equal SNR levels, for this reason each sub-channel can be modeled as a flat fading channel with a given SNR.

Decode-and-Forward (DF) scheme is used at the relay nodes where the signal is demodulated, decoded, re encoded and finally the signal is forwarded which was received from the source terminal during the first phase. Repetition based relaying, where the relay repeats the information received from the BS is considered. Hybrid-Automatic Repeat Request (ARQ) is provided by using transmit and receive diversity scheme. Also in the model the MAC-Protocol Data Unit (PDU) packets are transmitted in Forward Error Correction (FEC) blocks [14].The receivers use cyclic redundancy check to check whether a block is received correctly or not. The probability of not detecting a block is assumed to be negligible. Even if one bit is received in error the block is discarded. ARQ is not implemented. AMC is then used for each sub-channel and frame based on the selection method and low complexity end-to-end link adaptation. The considered modulation modes are BPSK, QPSK, 16-QAM and 64-QAM. The considered FEC includes convolutional coding with the following code rates: 1/2, 2/3, 3/4, 5/6, 7/8 and 1 [15]. Combining each modulation and coding modes gives one AMC mode. Since AMC is used, keep the transmit power from the relays and the BS constant. The terminals in the network consist of single antenna.

Throughput can be defined as the number of payload bits per second per hertz and per channel used and which are received correctly at the receiver.

### 3.3.2. Cooperative diversity schemes

For all the cooperative diversity schemes considered, the transmission for each user in each phase occurs at a given sub-channel j. The various cooperative diversity schemes are:

Cooperative Transmit Diversity–1

The MS (main station) and RS (relay station) listen to the transmission of the BS during the first phase. In the second phase, both BS (base station) and RS transmit simultaneously to the MS. When the same AMC mode is used for the two phases such transmission scheme can be realized, provided that the two phases have equal duration. Base station and Relay station uses cooperative space time coding in the form of Alamouti scheme [16].

Cooperative Transmit Diversity–2

Cooperative diversity schemes 1 and 2 are almost same; the difference is that, in cooperative transmit diversity-2 the main station does not modifies the signal received during the first phase in any form. Therefore, the AMC mode in each phase can be chosen independently and the two phases do not have to have equal duration.

Cooperative Receive Diversity

It also consists two phases, out of which in the first phase the source transmits at a particular AMC mode while the relay and the destination nodes receive the signal. In the second phase, the relay repeats with the same AMC mode and the BS remains silent. After Maximum Ratio Combining (MRC), the MS achieves cooperative receive diversity. Even if this scheme can achieve the same post processing SNR as that of cooperative transmit diversity–2, it suffers from a potentially higher multiplexing loss due to the need for identical AMC modes and hence equal–duration phases. Hence, cooperative receive diversity cannot outperform cooperative transmit diversity–2.

Cooperative Selection Diversity

With conventional relaying, the S → R transmissions occur in the first phase. The destination chooses not to receive during the first phase. In the second phase, only the relay transmits. The destination relies solely on the signals received via the R → D link. Base station chooses between conventional relaying and direct transmission in this scheme.

Adaptive Cooperative Diversity Scheme

Adaptive cooperative diversity scheme chooses the best scheme (in terms of end-to-end throughput) among direct transmission and the aforementioned cooperative diversity schemes. If the two schemes have the same performance the one with less complexity is selected. One has to order the schemes with increasing complexity as follows: direct transmission, conventional relaying, cooperative transmit diversity–2 and cooperative transmit diversity–1 [17]. This adaptive scheme chooses coherent signal combining at the MS only when it can increase the end-to-end throughput as compared to both conventional relaying and w/o relay schemes. Hence, it can reduce the complexity at the receiver while maximizing the end-to-end throughput.

### 3.3.3. The frame structure

Figure 14 depicts the frame structure for low mobility users in two-hop cellular networks with infrastructure based relays. In the figure, cooperative selection diversity based transmissions are considered. The users report their CSI to the BS using the fast feedback channel. Based on this CSI, the BS allocates the radio resources and schedules the users. It transmits in DL-MAP the information on which user is scheduled on which sub-channel and for each scheduled user whether relaying should be used or not. All the relays and users listen to this information. The duration of the second phase is fixed. If conventional relaying is selected for a given user, then the duration of the first phase at each sub-channel can be variable depending on the AMC mode chosen for the second phase.

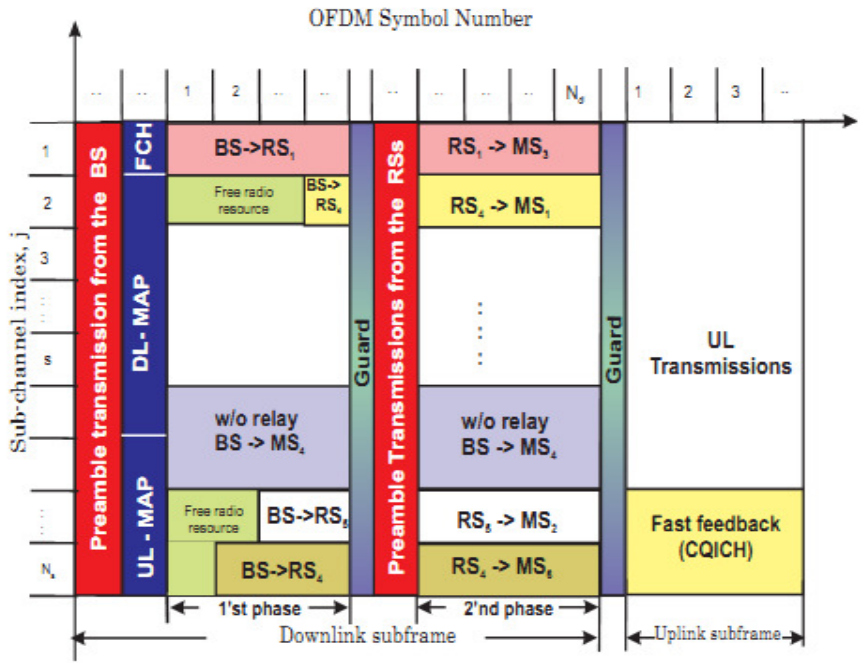

Figure 14.Frame structure for low mobility users in two-hop cellular networks with infrastructure based relays.

### 3.3.4. Relative Performance Evaluation of the Cooperative Diversity Schemes

Performance evaluations are done using the scheduling and radio resource allocation. The average end-to-end throughput is calculated per channel use, i.e., the average is taken over the radio resources allocated to the users in order to provide conclusions that are not sensitive to the system parameters. An FEC block is comprised of 96 coded bits. One sub-channel is comprised of 8 data sub-carriers and one pilot subcarrier over t consecutive OFDM symbols. The term t, t ∈ {2, 3, 6, 12} represents the number of OFDM symbols required to transmit one FEC block. It depends on the selected modulation mode with AMC. The duration of the second-phase is fixed to 12 OFDM symbols. The first phase can use up to 12 OFDM symbols. The scalable OFDMA mode with 1024 subcarriers with a system bandwidth of 10 MHz is considered. Consider users with speeds up to 7.7 km/h such that the 50% coherence time is greater than or equal to 10ms. The frames have 5 ms of duration. Time constant T is set to 100 in order to provide fairness to users. For the S → R links the wireless channel model is used with a path-loss exponent of 3 and a Rician factor K of 10. The selected model has a 90% coherence bandwidth of 17 sub-carriers. For the R → D and S → D links the Non-LOS (NLOS) channel is used with a path-loss exponent of 3.5. The total Effective Isotropic Radiated Power (EIRP) from the BS is fixed as 57.3 dBm. Since the relay terminals are simpler than a BS and transmit at lower power, we assume that the total EIRP from each relay station is fixed as 47.3 dBm. The heights of the MSs, BS and RS are 1.5 m, 32 m and 10 m, respectively. Carrier frequency is 2.5 GHz. Based on these assumptions; path–loss at each link is calculated accordingly. The effect of shadowing is not considered.

Our performance measures are the overall average throughput per channel use and the throughput gain for a single user at different positions in the cell. Average throughput gain of scheme A with respect to scheme B is defined as

$$\text{throughput}_{gain(AB)} = \frac{(\rho^A - \rho^B)}{\rho^B} \times 100$$

(13)

where $\rho^A$ and $\rho^B$ are the average throughput values of a single user at a given position in the absence of other users. Figure 15 shows throughput gain (cooperative transmit diversity-2, without relay). Also observe that around the BS, i.e., up to 6 km, the gain is zero. In this region direct transmission provides the highest end-to-end throughput.

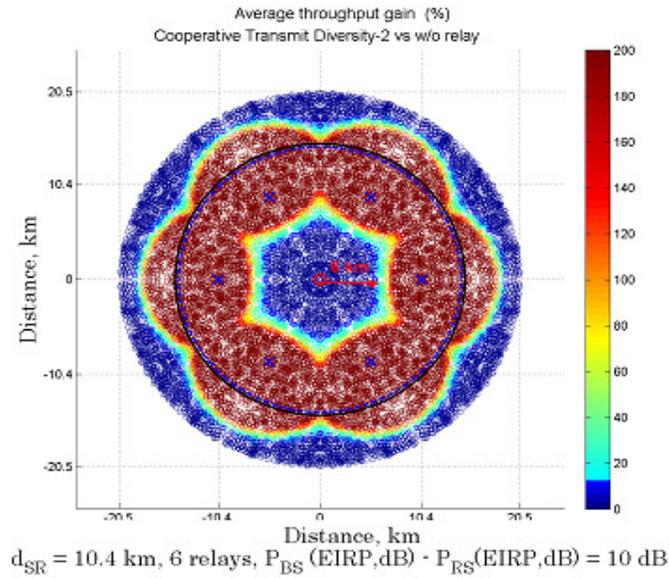

$d_{SR} = 10.4$ km, 6 relays, $P_{BS}$ (EIRP,dB) - $P_{RS}$(EIRP,dB) = 10 dB

Figure 15. Average throughput gain throughput gain (cooperative transmit diversity-2, w/o relay) at different position in the cell.

In the presence of multiple users in the cell, Figure 16 presents the overall average throughput per channel use as a function of the total number of relays in the cell. The average throughput is calculated within a range greater than 6 km and smaller than 14.85 km to the BS, which is the coverage area where relaying improves the performance compared to w/o relay system.

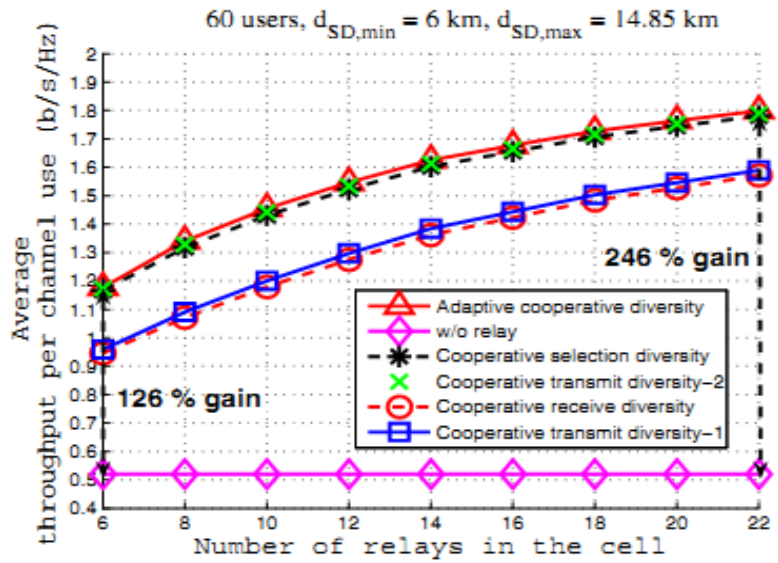

Figure 16. Average throughput per channel use versus the number of relay stations in the cell. Minimum and maximum distances of the users to the base station are 6km and 14.85km respectively

In Figure 17 is plotted the throughput gain (cooperative transmit diversity-2, cooperative selection diversity). The cooperative transmit diversity–2 brings a throughput gain of around 25% in most of the region where throughput gain is significant.

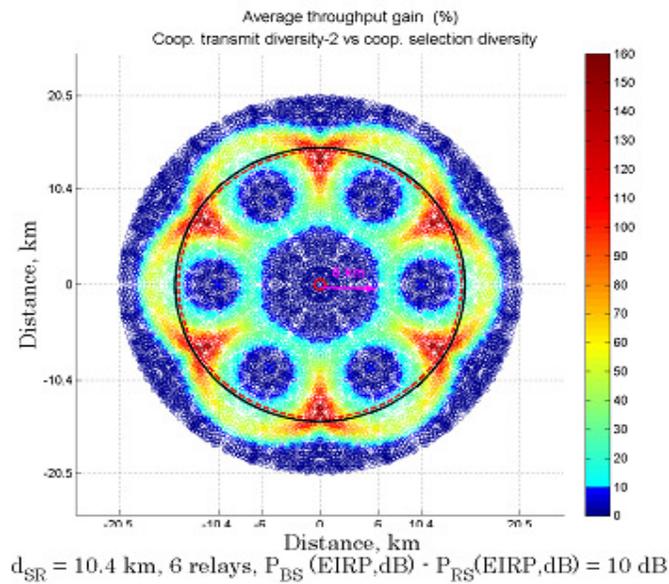

Figure 17. Average throughput gain throughput gain (cooperative transmit diversity-2, cooperative selection diversity) at different position in the cell.

Figure 18 shows throughput gain (cooperative transmit diversity-2, cooperative transmit diversity-1). The cooperative transmit diversity–1 can outperform the cooperative transmit diversity– 2 only at distances far from both the BS and the closest RS.

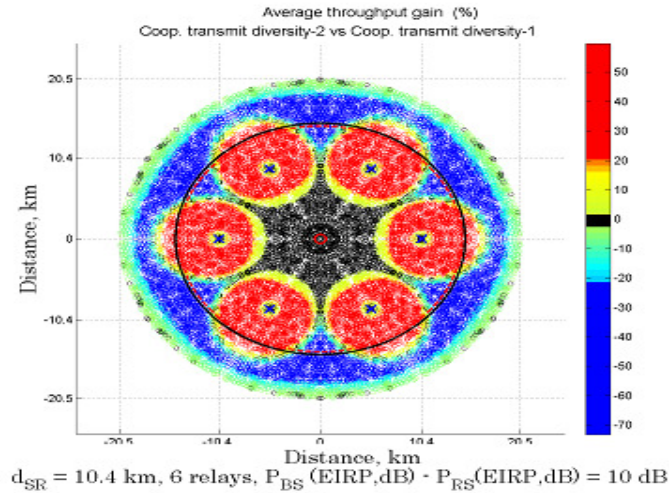

$d_{SR}$ = 10.4 km, 6 relays, $P_{BS}$(EIRP,dB) - $P_{RS}$(EIRP,dB) = 10 dB

Figure 18. Average throughput gain throughput gain (cooperative transmit diversity-2, cooperative transmit diversity-1) at different position in the cell.

## 4. CONCLUSIONS

In this paper, we have presented a theoretical analysis of cooperative diversity in various fields like wireless sensor networks, cognitive radio and resource allocation for IEEE 802.16j. For wireless sensor network analysis the knowledge of the spatial distribution of nodes is used to determine the number of packets to be transmitted as a function of distance from a sink. This number is a sum of packets due to MHR and due to cooperation. These numbers are then used in an energy analysis to determine the average energy used as a function of distance, thereby predicting network lifetime. In cognitive radio we have shown the benefits of cooperation in increasing the agility of cognitive radio networks. A simple two user cooperative cognitive network is first considered and showed to improve the agility by exploiting the inherent asymmetry. The cooperation scheme is then shown to be extended to multi-user multi-carrier networks. Lastly user scheduling techniques and various cooperative diversity schemes have been analyzed for resource allocation in IEEE 802.16j. Cooperative selection diversity scheme is shown to be a promising cooperative diversity scheme compared to the other more complex cooperative diversity schemes which require coherent signal combining at the mobile station.

**Authors**

Gurpreet kaur was born in India on February 25, 1987. She received her B. Tech degree in Electronics and Communication from University Institute of Engineering and Technology, Kanpur University, India in 2009 and currently is a M. Tech (Signal Processing) student in Mody Institute of Technology and Science (Deemed University), Rajasthan, India.

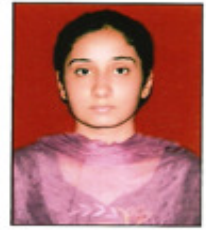

Partha Pratim Bhattacharya was born in India on January 3, 1971. He received M. Sc in Electronic Science from Calcutta University, India in 1994, M. Tech from Burdwan University, India in 1997 and Ph.D (Engg) from Jadavpur University, India in 2007.

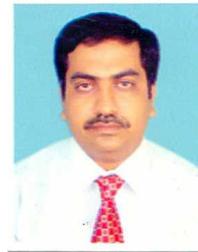

He has 15 years of experience in teaching and research. He served many reputed educational Institutes in India in various positions starting from Lecturer to Professor and Principal. At present he is working as Professor in Department of Electronics and Communication Engineering in the Faculty of Engineering and Technology, Mody Institute of Technology and Science (Deemed University), Rajasthan, India. He worked on Microwave devices and systems and mobile cellular communication systems. He has published a good number of papers in refereed journals and conferences. His present research interest includes mobile communication and cognitive radio.

Dr. Bhattacharya is a member of The Institution of Electronics and Telecommunication Engineers, India and The Institution of Engineers, India. He is the recipient of Young Scientist Award from International Union of Radio Science in 2005. He is working as the chief editor, editorial board member and reviewer in many reputed journals.